# Privacy-driven Access Control in Social Networks by Means of Automatic Semantic Annotation

Malik Imran-Daud[1], David Sánchez and Alexandre Viejo

*UNESCO Chair in Data Privacy, Department of Computer Science and Mathematics, Universitat Rovira i Virgili, Avda. Països Catalans, 26, 43007 Tarragona, Spain*

In online social networks (OSN), users quite usually disclose sensitive information about themselves by publishing messages. At the same time, they are (in many cases) unable to properly manage the access to this sensitive information due to the following issues: i) the rigidness of the access control mechanism implemented by the OSN, and ii) many users lack of technical knowledge about data privacy and access control. To tackle these limitations, in this paper, we propose a dynamic, transparent and privacy-driven access control mechanism for textual messages published in OSNs. The notion of privacy-driven is achieved by analyzing the semantics of the messages to be published and, according to that, assessing the degree of sensitiveness of their contents. For this purpose, the proposed system relies on an automatic semantic annotation mechanism that, by using knowledge bases and linguistic tools, is able to associate a meaning to the information to be published. By means of this annotation, our mechanism automatically detects the information that is sensitive according to the privacy requirements of the publisher of data, with regard to the type of reader that may access such data. Finally, our access control mechanism automatically creates sanitized versions of the users' publications according to the type of reader that accesses them. As a result, our proposal, which can be integrated in already existing social networks, provides an automatic, seamless and content-driven protection of user publications, which are coherent with her privacy requirements and the type of readers that access them. Complementary to the system design, we also discuss the feasibility of the system by illustrating it through a real example and evaluate its accuracy and effectiveness over standard approaches.

## 1. Introduction

Online social networks (OSN) such as Twitter, Facebook, Google+, Myspace, etc., are platforms where people interact with each other by publishing messages. In these platforms, users can build their own social circles of friends and join social groups or communities. In these groups and communities, strangers may connect with each other according to their common interests, views or activities [1]. In social networks, users spend most of their time in publishing or accessing information about such activities. Very frequently, the published content may contain sensitive data such as date of birth, political views, religious views, medical-related information or others.

Publicly shared content containing that sensitive information can be easily revealed by means of messages, profile data or social apps (like games). This data may portray a person's social or inner life [2], which constitutes a serious privacy issue. Social networks such as Facebook[2] and Twitter[3] consider trusted and non-trusted users as friends [3], but the trust of such friends cannot be measured [1]. As a result, sensitive information may be revealed to non-trusted users. Moreover, Johnson's analysis [4] concludes that the majority of users are more concerned with internal threats of privacy (i.e., from friends) rather than strangers. For this reason, most of the OSN friends are considered untrustworthy to share sensitive information. On the other hand, according to the *European Union Agency for Fundamental Rights & Council of Europe* [5], this sensitive information needs to be protected from untrusted third parties, because it can be exploited by such parties for their own benefit [6].

---

[1] Corresponding author. Address: Departament d'Enginyeria Informàtica i Matemàtiques. Universitat Rovira i Virgili. Avda. Països Catalans, 26. 43007. Tarragona. Spain
Tel.: +34 977 558270; Fax: +34 977 559710;
E-mail: malikimran.daud@urv.cat

[2] http://www.facebook.com
[3] https://twitter.com/

In the last few years, OSNs (such as Facebook) introduced some measures to improve users' privacy by implementing access control features. In order to incorporate such access control, the user profile is broken down into small customizable elements [7]. In order to manage the access to related resources, the information can be classified as "public", "private", "friend" or "friend of friend" [8]. According to Aïmeur et al. [7], these features are unreliable or fail to provide desirable results, because they are not fully understood [9] or it is difficult for the users to manage them correctly [4]. Furthermore, while configuring privacy settings, users need to perform a tedious job of defining policies for each user, type of resource and to classify those resources.

In order to overcome these shortcomings, the scientific community has proposed some access control solutions (Masoumzadeh et al [10] & Carminati et al. [11, 12]) that take into consideration the type of resources to be protected (e.g., photos, videos, wall messages, etc.) before allowing/rejecting an access request. These methods rely on ad-hoc structures (i.e., application ontologies) that provide a preliminary modeling of the resources. In order to manage the access control, the users or the OSN administrators need to define access control rules for each resource type. The proposed solutions bear some limitations. On one hand, the classification of resources is coarse grained, fixed and rigid. Similarly, access control policies are applied as a whole on the object or resource, regardless of their actual contents or sensitiveness. As a result, the access to the resource is binary, that is, complete access or complete restriction. For example, if a user declares *WallMessages* as private for a special group of friends, all the published messages will be hidden from that category of friends, regardless the messages contain any sensitive information or not. Furthermore, it is usually difficult for the users to configure the access control policies, since they may not be familiar with such notations and privacy issues.

In order to address the limitations introduced above, in this paper, we present a new scheme to enforce access control over resources published in social networks. We next summarize the main contributions of our work:

- We propose a transparent, dynamic and privacy-driven access control mechanism. Privacy is ensured by automatically protecting the content of messages to be published according to the privacy requirements of the publishers. The privacy requirements are defined by stating the type of information and the level of detail that is allowed to be accessed by each type of publisher's contact within the OSN. Contrary to access control policies defined over specific resources, such requirements are only defined once in a generic way and can be intuitively stated. Moreover, the user does not need to have a priori privacy notions.

- Contrary to related works, the privacy assessment is performed by semantically analyzing the contents to be published in an automatic way. Moreover, instead of evaluating the privacy for a resource (e.g., a publication) as a whole, our approach examines the privacy risk of each part of the resource individually (i.e., each textual term in a message).

- The semantics that drive the privacy assessment are gathered by means of an automatic semantic annotation process, which relies on available knowledge bases (i.e., DBPedia[4]) and several linguistic tools.

- In contrast to the binary access control policies proposed by other researchers (which just completely allow or deny the access to a resource), our access control enforcement provides each type of reader with a sanitized version of the original publication that is coherent with the privacy requirements specified by the publisher for that type of reader. The different sanitized versions are semantically coherent with regard to the original publication, and are created automatically according to the semantic annotation process and the privacy risk assessment.

The rest of this paper is organized as follows. In section 2 we review the available literature on this topic. In section 3 we present our access control mechanism and give a detailed description of its different components and how potential policy conflicts are managed. Section 4 illustrates the feasibility of the proposal through a real example. In section 5 we evaluate the system, under the perspectives of feasibility, scalability and accuracy of the privacy protection. Finally, in section 6 we provide some conclusions and present some lines of future work.

---

[4] http://dbpedia.org/About

## 2. Related Work

As introduced in the introduction, OSNs incorporate limited access control features in order to manage sensitive publications. More specifically, Facebook incorporates an option to split a list of friends into different categories, which are family friends, close friends, OSN groups or within the customized list of friends [4]. As a result, a user can specify the allowed categories before sharing her publications. However, such efforts are not practical enough because of the following reasons: i) the burden of configuring this access control for each publication, which requires knowledge about the privacy risks inherent to the publications and ii) the lack of flexibility of the system, because it is either granted or forbidden access to each publication. In fact, according to the survey conducted by Liu et al [13], only a 37% of the users are satisfied with this kind of privacy settings.

Researchers have also contributed to enhance the privacy of the users in OSNs. As a result, Carminati et al. [11, 12] categorized users and resources in an ad-hoc ontology in order to annotate OSN-related publications and modeled their relationships. In both schemes, the access control is enforced according to the relationships modeled in the ontology, and it is based on the trust level, type and depth of the relationships within users. Similarly, in Masoumzadeh et al. [10], the authors proposed a social network ontology to categorize different types of resources (e.g., photos, messages, etc.). Moreover, an access control ontology was also proposed in order to model the access control policy rules. In other related schemes, Pang et al. [14, 15] modeled users, their social relationships and their publically shared information (e.g., profile data and their publications) within different graphs. In this scheme, the access control is managed by means of policies that contain constraints and access rules for target users on the content that is publically shared by the owner. Access control is thus enforced by following the interrelationships of the users and their shared content within the graphs, and access decisions are taken according to rules defined in the policy. The solutions proposed by these authors have some common limitations. They are not flexible to modifications in the ontology, profiles or other contents, because they should undergo with a lot of manual changes in the ontology and also in the annotation of resources. Moreover, there is no mechanism defined to evaluate the sensitiveness of the resources, which leads the system to provide a coarse grained access. Therefore, the access to the resource is binary, that is, complete access or complete restriction. Besides, a lot of manual management by the users and the social network administrator is required in order to configure policies for each type of resource.

In another scheme, Cheng et al. [8] proposed a relational access control model, which is based on the concept of user to user and user to resource-based relationships. They proposed a regular expression-based language in order to specify access control policy rules. Moreover, they developed a path checking algorithm to determine relationships among the users and the resources from a social graph. In this solution, access is granted based on the relational policy defined among users and resources. As a result, access control can only be enforced if there is a direct relation of users and resources. Therefore, users need to perform a tedious job of defining a policy rule for each type of user and resource. In Villata et al. [16], the authors propose an access control mechanism based on a predefined vocabulary of sensitive terms. A user can customize her vocabulary of sensitive information and define access privileges for each reader. This mechanism is based on the ISICIL[5] social platform. This scheme improves (a specific aspect of) the models explained in previous paragraph, as it manages the access by semantically analyzing the text in order to find sensitive information. However, this scheme is static, as it only protects the limited information defined in the vocabulary, which includes profile data and some other information, such as time and place. Furthermore, the publisher has the burden of defining these vocabularies and policy rules for each user. Finally, Shehab et al. [3] proposed an access control mechanism in order to protect user's profile data from OSN applications (like games). In this mechanism, each user specifies profile attributes (e.g., date of birth, religion, etc.) that needs to be shared with OSN applications, and defines a level of access for each attribute. In this mechanism, a user manually configures profile attributes for each OSN application and, thus, the privacy protection is limited to profile data.

Bourimi et al. [17] proposed an interesting approach within the context of the European project *Digital.Me* that is related to our scheme. In this approach, privacy recommendations are triggered for social network users by analyzing semantic information disclosed within the OSN. For this purpose, the semantic equivalence among contacts is detected from profile data, in order to tackle linkability issues. However, this approach only triggers recommendations and does not sanitize sensitive information automatically. Moreover, it relies on profile ontologies and only deals with the profile attributes of the users and does not consider the sensitive information that may appear in the users' publications.

---

[5] http://isicil.inria.fr/

Summarizing, most of the previous approaches do not consider the content of the published data and/or they requires a lot of manual effort to configure the access to the resources individually. In contrast, our system is able to automatically assess the semantics of data without relying on ad-hoc structures, but just on a general purpose knowledge base (which is provided by the system). Then, the sensitive information of user's publications is automatically detected according to the semantics of the information and the privacy requirements of the user for each type of reader (which are just defined once). Accordingly, in order to ensure the user's privacy, it automatically protects the sensitive information by generating sanitized versions of the publications that are provided to each type of reader. As a result, the privacy enforcement is transparent both to publishers and readers, thus requiring no administrative efforts at the publication time. In addition, the proposed access control policies are flexible enough to be integrated into any social network.

## 3. Our Proposal

As shown in figure 1, the actors involved in our system are the *publisher* of a message, the *reader* of that message and the *social network*, which provides the framework. The *publisher* is responsible for specifying her privacy requirements and to publish data in the *social network*. The *reader* is the one who initiates a request for accessing to the published content; as a result, he gets a sanitized version of the publication in coherence with the *publisher's* privacy requirements with regard to him. The *social network* is in charge of controlling the whole process by i) semantically annotating the messages submitted for publication and ii) enforcing *publisher*'s privacy by creating semantically-coherent *sanitized versions* of the published content according to the type of *reader*. To tackle these tasks, two components are incorporated, respectively, in the *social network*: the *annotator* and the *monitor*. The following workflow of the system is explained with regard to figure 1.

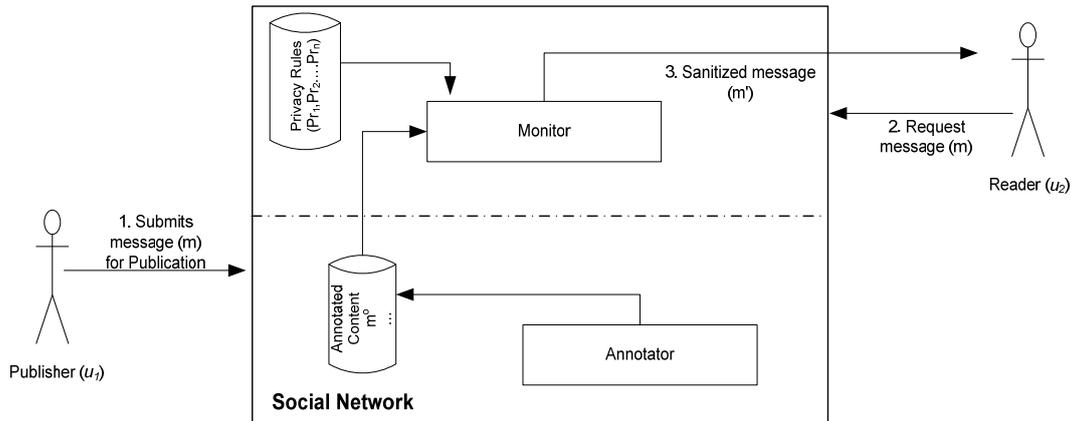

**Figure 1.** System architecture

In order to define her privacy requirements, the *publisher* specifies the level of content disclosure allowed for each type of contact in the *social network* in a generic way (e.g., only *family contacts* can know her *sexual orientation*). These requirements are stored as privacy rules (e.g., $Pr_1, Pr_2,......Pr_n$), which are associated to each *publisher* in the privacy rule database that is managed by the *social network*. The specification of the privacy requirements of each potential *publisher* is a process that is performed once, at the deployment stage of the system; afterwards, the system applies them for all of her future publications. The details of this process are explained in section 3.2.1.

Once the privacy requirements are specified, the subsequent workflow of the proposed system is as follows. For a message *m* to be published by the *publisher* $u_1$, the *annotator* module analyses the text of the message by performing syntactic and semantic analyses in order to identify and semantically annotate the content. The resulting annotated message $m^o$ is stored in the annotated content database within the *social network*. The annotation methodology of the *annotator* module is elaborated in section 3.1. After that, when a *reader* $u_2$ requests a message *m* of the *publisher* $u_1$, the request is evaluated by the *monitor* module. The *monitor* assesses the content sensitiveness of the annotated version of the requested message ($m^o$), with respect to the privacy requirements defined by the *publisher* $u_1$, which are retrieved from the privacy rules database. As a result, the *monitor* sanitizes the sensitive content of a message according to the level of disclosure allowed by the *publisher* ($u_1$) for the type of *reader's* contact (i.e., $u_2$) with respect to the *publisher*. The resulting sanitized message (m′) is finally forwarded to the *reader* $u_2$. The details of the *monitor* module are explained in section 3.2.

### 3.1. Submitting a Message for Publication

The *annotator* module is invoked whenever the *publisher* sends a message *m* to be published by the social network. The message contents are processed by the *annotator* module in order to perform the semantic annotation process. The formal semantics associated to the message content during the annotation are used in a later step to evaluate the sensitiveness of a publication, because our sensitiveness assessment is driven by the content of the message.

Within a discourse, nouns are the part of speech that provides the richest semantics and usually carry the sensitive content [18]. Therefore, the *annotator* module identifies the terms that are nouns from a given message, and then derives their semantics by associating them with a formal conceptualization. Since words may have multiple senses, there is a need to resolve the ambiguity by choosing the appropriate word sense that corresponds to the actual semantics of a message. Consequently, the *annotator* module also performs a word sense disambiguation to select the most appropriate sense in order to semantically annotate nouns. The activity diagram in figure 2 depicts the workflow of the *annotator* module that is explained in the following paragraphs.

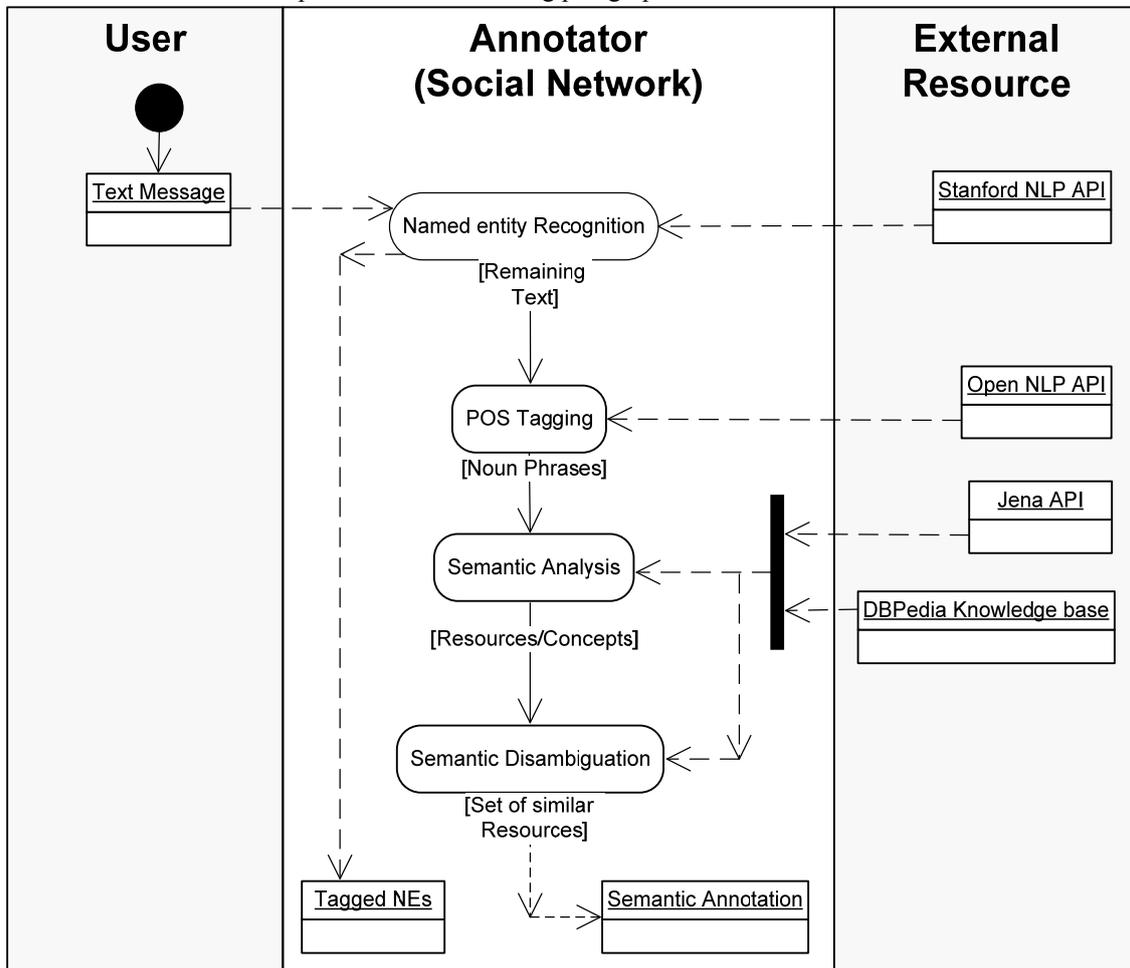

**Figure 2.** Semantic annotation activity diagram

Two different types of nouns that potentially carry sensitive data are distinguished: i) proper nouns (usually referred as *named entities* (NEs)), which are instances of concepts (e.g., person names, locations, etc.), and ii) common nouns that refer to concepts (e.g., a sensitive disease, a sexual orientation, etc.). The former are especially relevant from the privacy-preserving perspective, because their specificity and the fact that they usually identify an individual (e.g., a person) may produce a privacy leak. In order to identify them, we rely on a Named Entity Recognition package [19], which is able to identify named entities and classify them into seven categories: *Time, Location, Organization, Person, Money, Percent* and *Date*. Since this package relies on the lexical regularities of named entities (e.g., they are usually expressed with capital letters), rather than on the syntactical structure of a sentence, no priori analysis is needed to detect them. As a result of this process, the *annotator* module stores the tagged NEs and passes the remaining text of the message for further processing.

The second step of the annotation process consists on a *Part-of-Speech (POS) tagging* of the remaining text, which aims at identifying the common nouns potentially referring to sensitive topics (e.g., sensitive diseases). For this purpose, our system relies on a set of natural language processing libraries [20], which perform sentence detection, tokenization of terms, POS tagging and chunking. As a result of this step, noun phrases are identified.

In the next step, a *Semantic Analysis* is performed over the set of noun phrases provided by the *POS tagging* module. This consists in deriving the meanings of noun phrases by associating them with the concepts to which they refer in the message, which will serve as semantic annotation tags. To do so, our system relies on DBPedia [21], which provides a structured representation of Wikipedia articles (called resources) and which are taxonomically classified in the form of an ontology. Within DBPedia, the resources are classified into several knowledge bases such as the Wikipedia categories, YAGO and WordNet [21]. By exploring the classifications associated to a resource, we are able to associate a conceptualization to the noun phrases identified in the message that match with such resources.

In order to explore these classifications, we exploit the ontological properties of DBPedia resources that map them with each other based on their common categories. To do so, we use SPARQL [22] as a query language and the Semantic Web API [23] for its implementation. Accordingly, the SPARQL query is customized in the following order:

i. First it determines the DBPedia resources that contain a noun phrase identified in the message as a substring in its title (e.g., for a noun phrase like *Apple* we can have DBPedia resources such as *Apple, Apple_inc, Apple_iOS,* etc).
ii. Then, for each resource determined in step one; it derives a list of other resources that are linked with them on the basis of their relational properties (e.g., *Apple* ingredient_of *Fruit, Apple_inc* developer *Mac_OS,* where *ingredient_of* and *developer* are the properties and their ranges are other resources). As a result, an extended list of resources is gathered, which includes the resources that are semantically related to those retrieved in step-i by simple keyword matching. With this step, we aim to extend the number of possible conceptualizations of the noun phrase, which will be useful to derive the semantics of the noun phrase in the message and, thus, to annotate it.
iii. Finally, the system determines the Wikipedia categories of the resources gathered in step-ii (e.g., Wikipedia categories for *Apple* are: *Apples, Malus, Plants with sequenced genomes and Honey plants,* whereas, categories for *Apple_Inc* are: *Apple Inc, Computer companies of the United States, Computer hardware companies, Electronics companies, Steve Jobs and others*).

As a result of the *semantic analysis*, a set of possible conceptualizations (i.e., each one representing a word sense) and their taxonomic categories are retrieved for each noun phrase. If several conceptualizations have been retrieved, the most appropriate one (according to the sense) to which the noun phrase is referring within the message should be determined (e.g., for an ambiguous noun phrase like *Apple* we get conceptualizations like *Fruit* and *Apple_inc*). To perform the *Semantic Disambiguation*, our system uses the other noun phrases in the message as contextual information. Specifically, it calculates the semantic similarity of all senses retrieved for all noun phrases, and selects the combination of senses (one for each noun phrase) that are, in aggregate, the most semantically similar with each other. This strategy relies on the hypothesis that, to be semantically coherent, the senses of the noun phrases appearing in a sentence should refer to a common topic (i.e., they should be semantically similar). This is in fact the premise of most semantic disambiguation approaches proposed in the literature (developed around the senseval initiative [24]), which exploit the context of a term to identify its appropriate sense.

To compute the semantic similarity between senses, our system relies on the taxonomic structure of DBPedia. Because the semantic similarity between two concepts is understood as their degree of taxonomic resemblance [25] (e.g., *flu* and *pneumonia* are similar because both are respiratory disorders), semantic similarity is usually computed as a function of the commonalities and/or differences between their taxonomic ancestors. Specifically, in Sánchez et al. [26] the authors propose to measure the semantic distance (i.e., the inverse to similarity) between two concepts $a$ and $b$, as the ratio between the number of taxonomic ancestors ($T(a)$ and $T(b)$) that they do not share (as an indication of *distance*), divided by the union of both ancestors' sets (to normalize values to the size of the sets of ancestors). The logarithmic function is used as non-linear smoothing of the differences between the compared concepts (which better correlates with human assessment of similarities/distances), and the (1+) factor is added within the expression to avoid Log(0) calculations (i.e., in case of synonyms with identical sets of ancestors) and to ensure that the distance is within the [0..1] range [26]:

$$dist(a,b) = \log_2\left(1 + \frac{|T(a) \cup T(b) - T(a) \cap T(b)|}{|T(a) \cup T(b)|}\right) \qquad (1)$$

By applying this measure to all pairs of senses amongst all the possible combinations of senses retrieved from DBPedia for the noun phrases identified in the message, we can identify/disambiguate the most appropriate combination of senses as the one that, in aggregate, results in the smallest semantic distance (i.e., the highest similarity). As a result, each noun phrase is semantically annotated with the conceptualization associated to the disambiguated sense. The result of the semantic annotation is stored by the *annotation* module in the *annotated content database*. This semantic information will be the base for assessing term sensitiveness and performing the privacy driven access control in the next stages.

### 3.2. Accessing a Message

In this section, we propose an access control system for messages published in the OSN, where the policies are seamlessly defined by the owners of the resources, who are in charge of categorizing their OSN friends and specifying their allowed level of access to sensitive information. Note that the use of user categories is analogous to the use of roles in the well-known role-based access control (RBAC); therefore, our proposal can be considered to be inspired by the RBAC model but working in a discretional way [27].

Moreover, our system is flexible enough to be integrated in any OSN that implements publications, and it relies on existing OSN procedures in which an access request for a resource is monitored for authorization. To do so, our system requires a *monitor* module to be deployed in the OSN, which is responsible for the authorization of *reader's* access request for any given message. The *monitor* processes each access request based on the following three inputs: i) the annotated message of the publisher that a reader desires to access, ii) the reader classification with regards to the publisher and iii) the publisher's privacy requirements, which are defined by her to manage the access to her publications.

In order to process an access request, the *monitor* module retrieves the requested annotated message from the *annotated content* database. This message is annotated with three tags i) the publisher of a message, ii) the co-publisher of a message (i.e., the users who are tagged by the publisher in the message because the message content may refer to them), and iii) the semantic annotations automatically defined by the *annotator* module (discussed in section 3.1). The publisher and the co-publisher tags are provided by a user who publishes the message in the social network (these tags are currently supported and managed by most OSNs). On the other hand, the semantic elements are calculated and stored in the *annotated content* database by the *annotator* module.

The *monitor* processes an access request by annotating the reader according to the type of contact that he represents for the publisher. This contact type represents the nature of the relationship between the publisher and her contacts (e.g., *close friends*, *family friends*, etc.), which are defined by the publisher by following existing OSN settings and can be assigned while accepting a user's request for friendship. This categorization of friends reduces the administrative efforts of the users that define the privacy requirements for once on the whole group of friends, and the access is automatically managed for all future publications. Based on this reader annotation, the *monitor* assesses and manages an access request by determining the reader's type of contact with the publisher, and by applying an appropriate privacy rule that is defined by the publisher for that type of contact. These rules are defined according to the publisher's privacy requirements, which define the level of disclosure of information for each type of contact. According to such requirements, the system automatically generates related rules and stores them in a local repository of the social network. The definition of rules is a onetime process, that is, at the time of creation of an account, and the system automatically manages all future access requests to the publisher's messages according to these rules. The process required to retrieve the rules according to the privacy settings is detailed in the next section.

#### 3.2.1. Defining Access Rules According to Users' Privacy Requirements

An important goal of our system is to ensure the user's privacy while minimizing her administrative efforts to manage access to her publications. To do so, our system facilitates the users to configure their privacy requirements at the time of creation of OSN account and, then, it automatically and seamlessly manages the access to their publications according to these requirements. By means of these requirements, the system defines a list of rules that contains the access levels of disclosure to sensitive data for the publishers' types of contacts in the OSN. The following paragraphs elaborate the process of rules specification and management.

The rules are defined according to three types of elements: i) the sensitive topics (*ST*), that is, the type of data that are sensitive according to, for example, privacy regulations [28], ii) the contact categories (*CC*), which are defined by the publisher, and iii) the access level (*AL*), that is, the level of sensitive information disclosure allowed for a contact type. In our approach, the user can choose the sensitive topics that she wants to protect from others (e.g., religion, race, health, etc.) from a list provided by the system. On the other hand, the user manages her OSN contacts by classifying them into distinct categories (as discussed in section 3.2). This categorization of friends is based on the level of trust. After that, the user defines access levels by providing a list of terms that represent the maximum degree of information disclosure for each sensitive topic and each contact category (*CC*). By means of that, each user can control the access to her sensitive data by restricting the level of information detail that will be provided to each type of contact. The following tuple represents an access rule.

$rule_i \equiv \langle st_i, cc_i, al_i \rangle$

Where, a *rule$_i$* ∈ *Rules* has the following elements: i) a *sensitive_topic (st$_i$)*, ii) a *contact_category (cc$_i$)*, and iii) an *access_level (al$_i$)*. Sensitive topics *ST*={*st$_1$,st$_2$,..., st$_n$*} could be any topic that is considered sensitive. Contact categories *CC*={*cc$_1$, cc$_2$, ..., cc$_m$*} are those implemented by the OSN. Access levels *AL*= {*al$_1$, al$_2$,..., al$_m$*} are defined by terms representing the maximum level of information disclosure for each element in *CC*. In the following paragraphs we discuss each element in detail.

**Definition 1:** *Sensitive topics (ST)*: A set of topics that are provided by the system and may be sensitive (e.g., according to privacy legislations) because it portrays the information about a user that can be misused if disclosed to others.

For example, according to the *European parliament and the council of the EU* [28] and the U.S. laws on medical data privacy [29], sensitive individual's data is such that is related to *medical health, religion, race, politics* and *sexuality*. In view of that, the users can define their privacy rules related to these topics that are considered as sensitive topics by our system.

**Definition 2:** *Contact categories* (*CC*): The users classify their contacts into distinct categories based on the level of trust. Examples of *CC* can be *close friends, friends, family*, etc., as defined by the OSN.

This categorization is helpful in order to reduce the administrative efforts of the users, because, by defining a rule for each category of friends, the system can automatically manage the access to new members of this group or to users that can change between categories (e.g., a *friend* becomes a *close friend*). However, the possibility to define rules for specific individuals is also supported by the system.

**Definition 3:** *Access levels (AL)*: The publisher can restrict the access to the contents of her publications by defining the allowed level of disclosure of information that is related to a specific *ST*, and by assigning it to each type of contact in *CC*. Thus, terms in *AL* define the maximum level of information that a reader of a certain *CC* type can access in any message of the publisher.

*Example 1*: A user B*ob* configures his privacy settings related to *medical health* (that is *ST*), defines access levels *AL* by specifying terms *AL*={*Disease, Hepatitis*}, and classifies her OSN friends into the following OSN contact categories *CC*={*close friends, family friends*}. Hence, the access level *AL* assigned to *close friends* is *disease,* whereas, the level of access for *family friends* is *hepatitis*. As a result, any publication of *Bob* that contains information about *hepatitis* will not be completely disclosed to *close friends* (in fact, they will get a sanitized version of the message, as it will be explained in section 3.2.2). On the other hand, *family friends* will get the information about *hepatitis* but not more specific details (e.g., types of *hepatitis B, hepatitis C*). The following rules are generated as a result of these settings.

*rule$_1$*=<*medical health, close friends, 'diseases'*>
*rule$_2$*=<*medical health, family friends, 'hepatitis'*>

In the following examples, we illustrate the process of rules definition and their enforcement by the system.

$rule_3 \equiv\ <religion, friends, \text{'religion'}>$
$rule_4 \equiv\ <religion, family\ friends, \text{'Muslim'}>$

*Rule₃* restricts *friends* contacts to get any details of the publisher's *religion* (e.g., the publishers belief, sect, others), whereas, *rule₄* permits the *family friends* to know that the publisher is *Muslim*, but anything more specific will be sanitized.

$rule_5 \equiv\ <sexuality, friends, null>$
$rule_6 \equiv\ <sexuality, family\ friends, \text{'homosexual'}>$

In *rule₅*, the level of disclosure for *friends* is *null*, that is, *friends* contacts cannot get any information related to the sexual life of the publisher. As a result, any information related to the publisher's sexual life will be sanitized from any publication accessed by *friends*. In contrast, the *rule₆* permits *family friends* to know that the publisher is *homosexual* but nothing more detailed.

The previous rules are defined at a conceptual level, as a function of the semantic annotation performed by the *annotator* module. Since the annotation process is also able to detect and classify Named Entities (NEs), rules can be also defined in order to protect specific types of NEs (e.g., *persons*, *locations*, *organizations*, etc.) that, due to their specificity, may reveal sensitive information.

$rule_7 \equiv\ <NE\_person, strangers, null>$
$rule_8 \equiv\ <NE\_person, family\ friends, person\_name>$
$rule_9 \equiv\ <NE\_location, family\ friends, location\_name>$
$rule_{10} \equiv\ <NE\_organization, family\ friends, organization\_name>$

*Rule₇* restricts *strangers* contacts to get any information that refers to *person names*. However, *rule₈*, rule₉ and rule₁₀ permits only *family friends* contacts to access the specific name of a *person*, a *location* or an *organization* mentioned in the publications.

Notice that, according to the nature of the sensitive topics considered in the requirements, the protection will focus on confidential data (e.g., sensitive diseases, sexuality, etc.), thus protecting against *attribute disclosure*, or on identifying data (e.g., person names, locations etc.), thus protecting against *identity disclosure*.

### 3.2.2. Enforcing Flexible Access Control

As mentioned in section 3.2.1, publishers manage their privacy requirements by assigning a level of information disclosure to each contact type. To enforce the appropriate access to sensitive data according to such requirements, the publisher's messages are assessed for sensitiveness according to the type of reader that is accessing it. This sensitiveness is determined according to the following elements: i) the contact type of the reader, ii) their allowed level of disclosure, as defined in the rules, iii) the taxonomy associated to the access level (*AL*) term and iv) the semantic annotations of the message. The contact type of the reader is evaluated in coherence with the privacy requirements of the publisher, in which she categorized her friends in to distinct contact categories (*CC*). According to the contact type of the reader, the rule assigned to them is retrieved from the *privacy rule* repository (which is managed by the social network). As a result, the access level (*AL*) assigned to this contact type is determined from the rule assigned to him. The taxonomy associated to access level terms is retrieved from DBPedia. Finally, the annotated message of the publisher is retrieved from the *annotated content* database.

The system measures the sensitiveness of each term in the message by comparing their semantic annotations with the assigned level of access allowed for the reader. To do so, the taxonomic branch of assigned access level term is retrieved from DBPedia, in which the top level node is the access level (*AL*) term. As a result, any content that lies under the access level node is considered as sensitive for the reader and it is sanitized (i.e., replaced) with the term that is defined in *AL*, which defines the maximum level of disclosure for that type of reader.

*Example 2*: By considering *rule$_1$* and *rule$_2$* defined in *Example 1* (section 3.2.1), *Bob* publishes a certain text that contains the term *hepatitis*. Then, a contact named *Alice,* who is classified as a *close friend* of *Bob* tries to access that message. The *monitor* intercepts the request and it checks the rule assigned to the contact type *close friend* (i.e., *rule$_1$*) in order to determine the level of access for Alice (which is *diseases*). Then, it retrieves the taxonomy branch related to *diseases* (shown in figure 3) from DBPedia and checks that the term *hepatitis* lies under the *AL* (i.e., *diseases*). Finally, the monitor sanitizes (i.e., replaces) the term *hepatitis* by the *AL* (i.e., *disease*) allowed to *Alice* and provides this sanitized message to her.

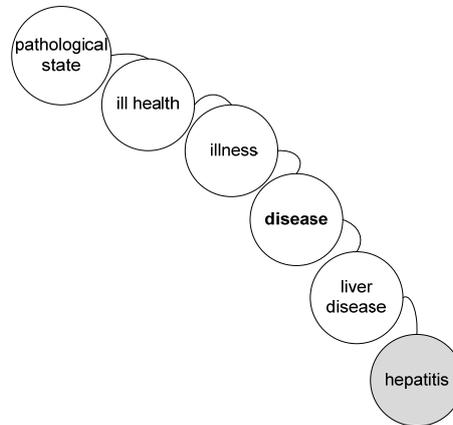

**Figure 3.** Taxonomic generalizations of Hepatitis

### 3.2.3. Policy Conflicts

In addition to the sanitization of messages, the *monitor* also handles potential *policy conflicts* within the users. A conflict may appear when the publisher posts a message on another user's timeline or tags another user in her publications, because the message content may refer to the latter. To handle the privacy requirements that may apply in such cases, in our system, the users (other than the publisher) who are associated with the content of a publication become the co-publishers of the publication and their privacy rules are also taken into consideration. However, the fact that several rules are associated to a message may cause a *policy conflict* between the publisher and the co-publisher(s), because the publisher and co-publisher(s) may define different access levels for their types of contacts (e.g., *close friends*, *friends*, *strangers*, etc.). As a result, there can be conflicting policies for their contact types. In this situation, in order to fulfill all the privacy requirements, the strictest rule amongst those of the publisher and the co-publisher(s) is the one that will be applied for their respective contact types. In practice, this means that the access level of disclosure that is higher in the taxonomic tree (i.e., the more generic in terms of semantics and, thus, the one that imposes the strictest restriction with regard to term sanitization) is the one that will be considered to sanitize the message contents.

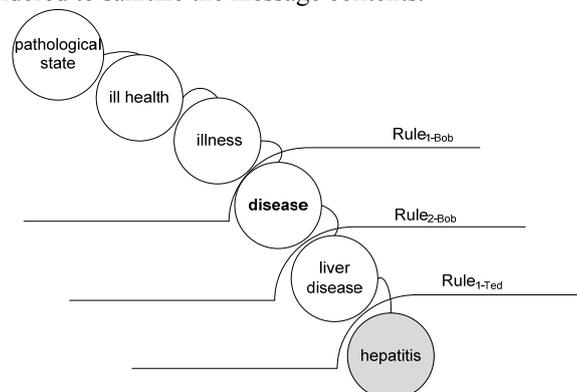

**Figure 4.** Access levels defined by Bob and Ted

*Example 3*: Extending *Example 1* and *Example 2* (section 3.2.1), *Bob* tags her friend *Ted* in a publication. As a result, *Ted* becomes co-publisher of the message. *Alice* can be a common friend or a stranger for *Bob* or *Ted*. The access levels defined by *Bob* and *Ted* are depicted in figure 4.

Accordingly, the following rules are generated as a result of privacy requirements of both users.

Rule$_{1\text{-Bob}}$ ≡ ⟨ *medical health, strangers, 'illness'* ⟩
Rule$_{2\text{-Bob}}$ ≡ ⟨ *medical health, close friends, 'disease'* ⟩
Rule$_{1\text{-Ted}}$ ≡ ⟨ *medical health, close friends, 'liver disease'* ⟩

In the following cases, we discuss the automatic resolution strategy for the policy conflicts:

**Case 1:** *Alice* is a *close friend* of *Bob* and *Ted*

The access request by *Alice* to *Bob*'s publication, with *Ted* as a co-publisher, produces a *policy conflict*; that is, there are two different levels of disclosure for *Alice* by the publisher and the co-publisher of the message (i.e., Rule$_{2\text{-Bob}}$ and Rule$_{1\text{-Ted}}$). In order to resolve the conflict so that the privacy requirements of both *Bob* and *Ted* are fulfilled, the *monitor* retrieves the taxonomic branches (show in figure 4) of both *ALs* defined by the publisher and co-publisher (i.e., *disease* and *liver disease*, respectively) and selects the level that is more general, that is, higher in the taxonomic tree. As a result, the level of disclosure *disease* is chosen by the *monitor* in order to sanitize the message content for *Alice*.

**Case 2:** *Alice* is a stranger to *Bob* and a *close friend* of *Ted*

In this case, the level of access for *Alice* defined by *Bob* is *illness*, whereas by *Ted* it is *liver disease*. By applying the same resolution strategy as above, the sanitization of the content of the message will be performed by using *illness* as access level.

## 4. Feasibility study

In this section, we illustrate the practical feasibility of our system by analyzing the scalability of its different modules, and illustrate its behavior with an example framed within Social Network specialized in healthcare.

As already discussed in section 3.2.1, it is also important to recall, that the specification of the privacy requirements (that would be done just once, during the initialization of their user accounts), is the only interaction that the system requires from the user. The rules generated as a result of these requirements and the automatic assessment of sensitive information that is driven by the semantic annotation will provide the means to enforce a transparent and automatic access control over all the subsequent publications. These privacy requirements can be based on the sensitive topics (*ST*) defined in current legislations on data privacy, such as the EU Data Protection Act [28] (i.e., *medical health, religion, race, politics and sexuality*), U.S. laws on medical data privacy [29] (which define lists of sensitive diseases such as *HIV, hepatitis, sexually transmitted diseases*, etc.) or the Health Insurance Portability and Accountability Act (HIPAA) [30] (which specifies the protection identifying census features such as *names, locations*, etc.). Furthermore, the topics can be chosen according to the thematic scope of the OSN. Finally, the access levels (*AL*) to be defined for each topic would match the number of contact categories (*CC*) in the OSN that is 3, in average, according to [11, 12]. Thus, the definition of privacy requirements requires minimal manual efforts by the users.

To illustrate this, let us compare this with the configuration burden of a standard approach in which the users would need to specify the access permission for each publication and contact type. To enforce the same level of access control that our system provides in a standard OSN, the user would need to i) assess the sensitiveness of the contents to be published for *every* new message, ii) for each message with sensitive contents, create as many sanitized versions of the message as contact types with different privacy requirements and iii) define the appropriate access control rules so that only the allowed contact types can access to the corresponding message. According to the Statistic Brain Research Institute [31] and kissmetrics [32], on average, a Facebook user publishes 90 pieces of content per month; from these, around a 58% of the publications require privacy-conscious settings [13]. Thus, users should manually evaluate those 90 pieces in order to identify which of them may cause privacy risks and protect the 52 (90*0.58) pieces that, in average, are sensitive. If we consider an average of 3 contact types, then the user would need to create 3*52 message versions and define 3*52 access rules, in the worst case. In comparison, in our approach the user only needs to specify as many access levels (*AL*) as contact types

(*CC*) per sensitive topic (*ST*). If we consider a generic implementation with 6 sensitive topics (those defined in the EU Data Protection Act plus census-related features), we have that the user only needs to define 6*3 *AL* just once, during the initial configuration step.

Let us now illustrate the whole process within the context of the sample social network. Because of the medical scope of this social network, the privacy protection can be restricted to *health*. Therefore, privacy rules can be configured so that they are related to the *health* topic. Let us also assume that contacts are categorized into the following three groups: *Clinicians/Researchers, Followers* and *Registered users*. Thus, for privacy rules, the access levels (*AL*) for the contact categories (*CC*) are related to the different levels of disclosure that may be allowed for the medical condition of the user of the SN (which is the sensitive topic (*ST*). The following sets show the customized access levels (*AL*) and the contact categories (*CC*) for this social network:

*AL* = {(HIV/AIDS/Hepatitis/STDs), Infections, ill health, Condition/State}
*CC* = {(Clinicians/Researcher), Followers, Registered users}

The first elements of *AL* (i.e., HIV/AIDS/Hepatitis/STDs) correspond to the diseases that are considered sensitive by the U.S. federal laws on data privacy [29], whereas, the rest of the elements are semantically coherent generalizations of the former according to the taxonomic structure of DBPedia.

Once the *AL* and *CC* are defined, the user can configure her privacy requirements by assigning a specific *AL* to each element in *CC*. To do so, the system provides an intuitive interface to define her privacy requirements in the form of questions related to the sensitive topics with a list of contact categories and a predefined set of taxonomically coherent access levels. Figure 5 shows an example of the list of *AL* that can be assigned to each element of *CC*.

**Figure 5.** Privacy requirements of a user in sample social network for the *medical health* topic

In this example, the access level for the *Clinicians/Researchers* group is *HIV/AIDS/Hepatitis/STDs*, whereas, the maximum allowed disclosure for *Followers* regarding these diseases is *infections* (but not specific diseases) and the *Registered users* are only allowed to know the general notion of *ill health*, but nothing more specific other than *ill health*. The formalization of these privacy requirements as rule tuples (rule$_i$ ≡ ⟨ $st_i$, $cc_i$, $al_i$⟩) are as follows:

rule$_1$ = ⟨ *Medical health, Clinicians/Researchers, HIV* ⟩
rule$_2$ = ⟨ *Medical health, Clinicians/Researchers, AIDS* ⟩
rule$_3$ = ⟨ *Medical health, Clinicians/Researchers, Hepatitis* ⟩
rule$_4$ = ⟨ *Medical health, Clinicians/Researchers, STDs* ⟩
rule$_5$ = ⟨ *Medical health, Followers, Infections* ⟩
rule$_6$ = ⟨ *Medical health, Registered users, ill health* ⟩

In order to show how these privacy rules are applied in practice, we use the sample message shown in figure 6, in which a patient shares her personal feelings about her disease (HIV). As per privacy requirements, this message is related to the *medical health* topic (i.e., *ST*) and is sensitive because it contains information about the disease of the publisher.

Dealing with Hiv and then being told that you suffer from AIDS is almost the hardest thing to face with in life. The hardest thing is dealing with the virus because there are people that just do not understand and think that you are a leper.
**Figure 6.** Sample message to be published

In order to process the publisher's messages, the *annotator* module performs the semantic annotation of the messages' content, as detailed in section 3.1. First, the message is syntactically analyzed to detect POS (see the output in figure 7 that corresponds to the sample message).

*Dealing with Hiv[NNP] and then being told that you suffer from AIDS[NNP] is almost the hardest thing[NN] to face with in life[NN]. The hardest thing[NN] is dealing with the virus[NN] because there are people[NNS] that just do not understand and think that you are a leper[NN].*

**Figure 7.** POS tagging of the sample message

Then, the semantic annotation is performed. Given the number of messages that an OSN is expected to receive on daily basis, the scalability of the annotation process is crucial. On the one hand, we can consider that the average length of publications in current social networks is relatively short and, in some cases, restricted by the number of characters (e.g., Twitter allows only 140 characters). On the other hand, our semantic analysis focuses on the noun phrases of the publication (marked as NN (singular noun), NNP (proper noun) and NNS (plural noun) in figure 7) and hence, it scales according to the cost of analyzing them. In order to perform the annotation, our system derives the semantics of noun phrases by finding their potential conceptualizations in DBPedia. For this purpose, SPARQL queries are performed to retrieve the possible senses and their corresponding taxonomic structures. Given that, the DBPedia queries are the most costly part of the annotation process, the cost of analyzing a message is proportional to the time ($T\acute{s}$) required to get the number of senses ($n_{senses}$) for each noun phrase within a given publication, which is formalized as follows.

$$Cost \propto \left\{ \sum_{i=1}^{n_{NP}} T\acute{s}(n_{senses}) \right\} \quad (2)$$

A single query is executed for each noun phrase to retrieve any number of senses from the knowledge base. Table 1 summarizes the actual runtime required to execute SPARQL queries with respect to the number of senses of different noun phrases of the message (as shown in figure 6).

| Noun Phrase | No of senses | Time (ms) |
|---|---|---|
| HIV | 5 | 0.24 |
| AIDS | 12 | 0.21 |
| Thing | 54 | 0.24 |
| Life | 86 | 0.33 |
| Virus | 100 | 0.19 |
| People | 12 | 0.21 |
| Leper | 86 | 0.33 |

**Table 1.** Time required by a query to get senses of noun phrases

We can see that the runtime of a SPARQL query is independent of the number of senses of each noun phrase, and that the average cost per noun phrase is 0.25ms. In consequence, as we can see in figure 8 that the runtime of the annotation is linear with regard to the number of noun phrases, that is, it scales linearly according to the number and length of the messages to be analyzed.

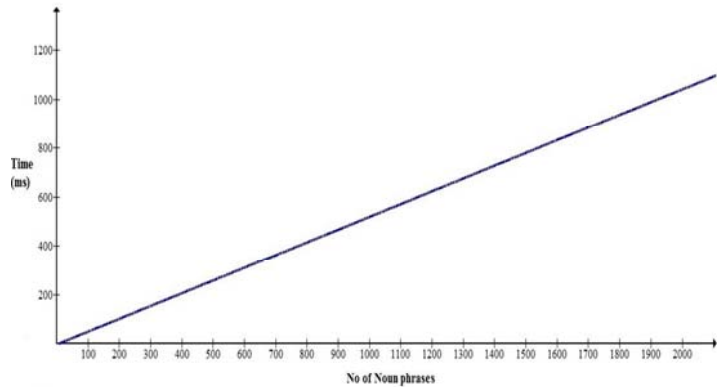

**Figure 8.** Runtime of the annotation process w.r.t. the number of noun phrases to analyze

As a result of SPARQL queries, a number of potential senses are retrieved for each noun phrase. Table 2 shows a list of senses of the noun phrases of the sample message.

| HIV | LIFE | VIRUS | AIDS | PEOPLE | LEPER |
|---|---|---|---|---|---|
| • Sexually_transmitted_diseases (STDs) | • Biological_science | • Viruses | • HIV/AIDS | • Humans | • Tropical_diseases |
| • HIV | • Systems | • Virology | • Pandemics | • People_(magazine) | • Leprosy |
| • Lentiviruses | • Biology | • Pediatrics | • Health_disasters | | • Bacterial_diseases |
| | • Life | • Organism | • Syndromes | | • Neglected_diseases |
| | | • Others | | | |

**Table 2.** Senses of the noun phrases in the sample message

As detailed in section 3.1, senses need to be semantically disambiguated in order to get the appropriate set of senses to get actual semantics of the message. To do so, our system calculates the semantic distance of each sense with respect to senses of the other noun phrases according to their taxonomic structure in DBpedia (that is retrieved as a result of the previous SPARQL queries). For example, let us consider *HIV* and *LIFE* nouns appearing in the sample message. The semantic distances between their senses are shown in table 3 that are calculated according to the taxonomies retrieved from DBPedia (which are shown in figure 9).

|     |      | LIFE  |        |         |                    |
| --- | ---- | ----- | ------ | ------- | ------------------ |
|     |      | Life  | System | Biology | Biological_science |
| HIV | STDs | 0.888 | 0.937  | 0.94    | 0.93               |
|     | HIV  | 0.857 | 0.928  | 0.93    | 0.92               |

**Table 3**. Semantic distances between the senses of *HIV* and *LIFE*

By performing the same process for all the senses of all the noun phrases, we obtain the set of most suitable senses, which are shown in table 4. Notice that the disambiguation process does not require any additional SPARQL query, but just the pairwise evaluation of already retrieved taxonomies.

| **HIV** | **LIFE** | **VIRUS** | **AIDS** | **PEOPLE** | **LEPER** |
| ------- | -------- | --------- | -------- | ---------- | --------- |
| *HIV*   | *life*   | *Viruses* | *AIDS*   | *Humans*   | *Leprosy* |

**Table 4**. Set of most suitable senses/annotation for the sample message

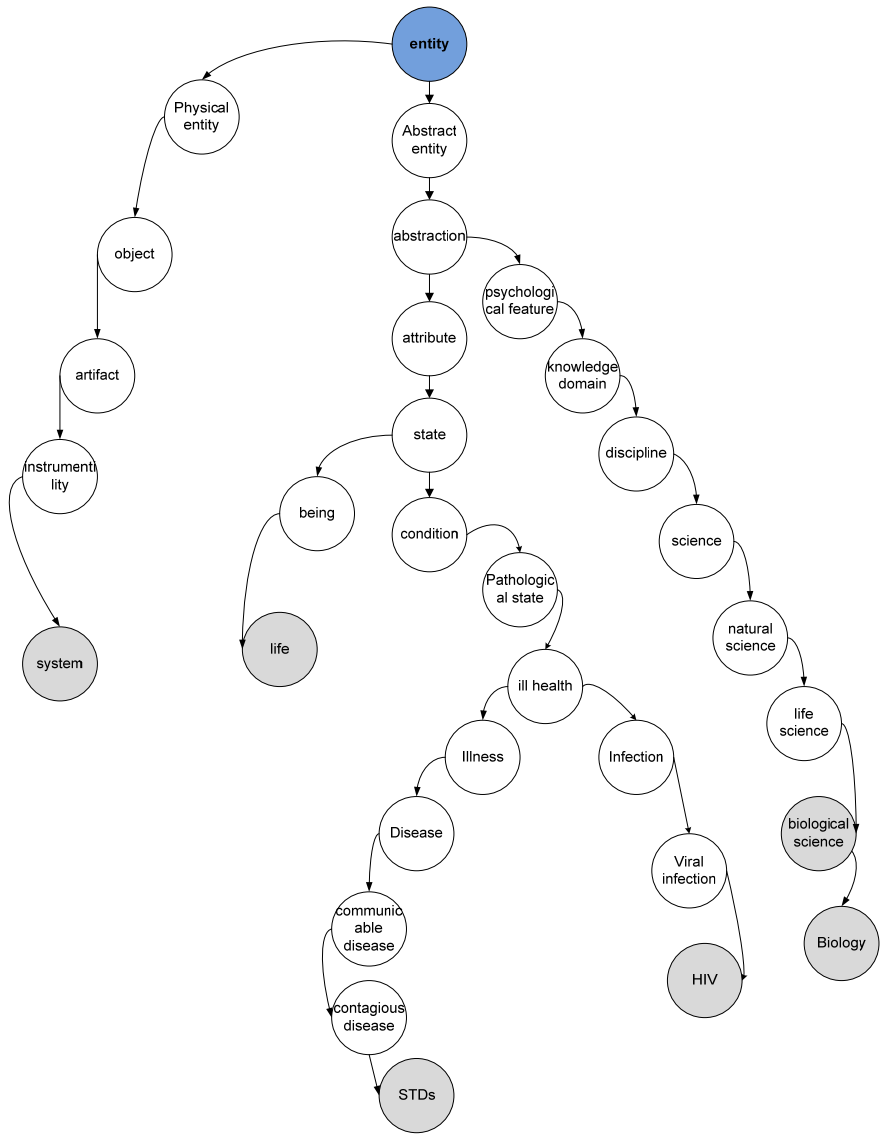

**Figure 9.** Taxonomic tree of senses for the sample message

Once the message is annotated and stored, the *monitor* processes access requests of the readers and assesses the sensitiveness of the messages according to the privacy requirements of the publisher (shown in figure 5). Eventually, any sense (shown in table 4) that lies within the branch of access level (that is, a level of disclosure for the contact type of the reader) is considered sensitive, and the corresponding noun phrase is sanitized accordingly. In this case, the computational cost for the sensitiveness assessment of a message for a specific contact type is proportional to the product of number of senses/annotations ($n_{senses}$) and the number of nodes ($r$) in a taxonomic branch of the access level corresponding to the contact type:

$$Cost \propto n_{senses} \times r \qquad (3)$$

Given the short length of typical messages, the limited amount of contact types and the fact that the taxonomies corresponding to the noun phrases have been already retrieved during the annotation stage, the sanitization process is highly scalable. Moreover, the numbers of distinct sanitized versions of a message are also limited to the types of contacts and access levels defined by the user and, thus, once they are created, they can be cached for further access requests by the readers of same contact type.

Let us illustrate this process for the sample message with respect to the privacy requirements (defined in figure 5). The sanitization process will be based on the taxonomies retrieved (listed in table 4) from DBPedia for each annotated noun phrase and the access levels defined for each contact type, which are shown in figure 10.

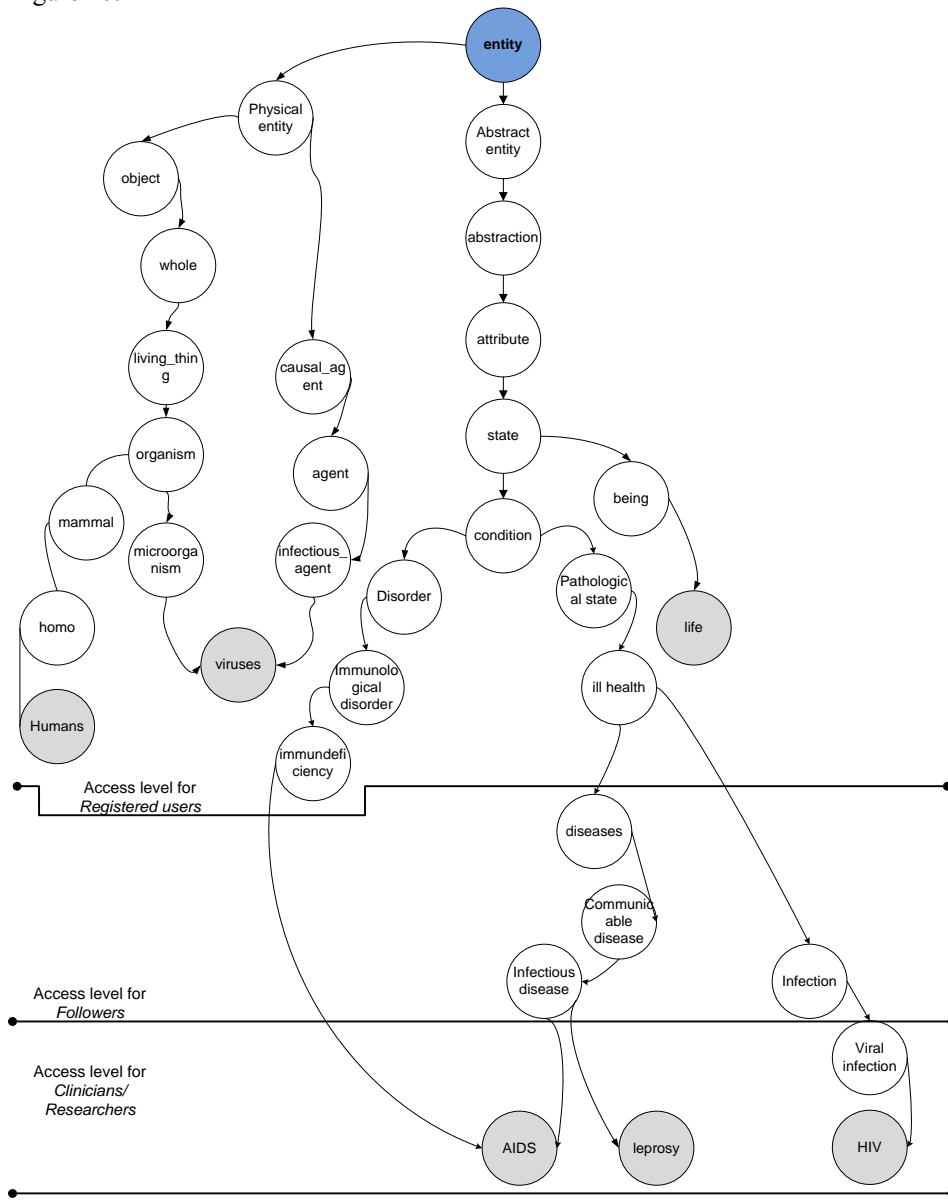

**Figure 10.** DBPedia taxonomies and access levels of the noun phrases in the sample message

According to rule$_5$, the access level for *followers* is *infections*. Therefore, the concepts below *infection* and *infectious disease* in the taxonomy shown in figure 10 (e.g., *AIDS, Leprosy* and *HIV*) are considered sensitive for *followers*. Therefore, the corresponding noun phrases need to be sanitized in order to fulfill the privacy requirements of the publisher. For this purpose, the sensitive noun phrases are replaced with the terms of access level nodes (i.e., *infection* and *infectious disease*). As a result, the *monitor* of our system will prepare a sanitized version of a message (shown in figure 11) for *followers*, which hides the specific details of the publisher's condition.

> Dealing with **infection** and then being told that you suffer from **infectious disease** is almost the hardest thing to face with in life. The hardest thing is dealing with the virus because there are people that just do not understand and think that you are a **infectious disease**.

**Figure 11.** Sanitized message for *followers*

Likewise, according to rule$_6$, the access level for *registered users* is *ill health*, thus producing the sanitized message shown in figure 12.

> Dealing with **ill health** and then being told that you suffer from **ill health** is almost the hardest thing to face with in life. The hardest thing is dealing with the virus because there are people that just do not understand and think that you are a **ill health**.

**Figure 12.** Sanitized message for *registered users* group

Finally, according to the first four rules (i.e., rule$_1$ to rule$_4$), the *clinicians/researchers* can access all the details of diseases of the publisher. Consequently, there are no sensitive terms that fall under the access level defined for this type of group. Therefore, the information will not be sanitized and the members of this group will get the plain text as shown in figure 6.

## 5. Evaluation

To complement the feasibility study that mainly considered the scalability of the system from both manual and algorithmic sides, in this section we evaluate the accuracy of the semantic annotation and the subsequent sensitivity assessment and privacy protection. To do so, we measure and evaluate i) the accuracy of the detection of sensitive terms; and ii) the accuracy of the semantic disambiguation process.

As evaluation data, we considered a set of entities related to the sensitive topics which are covered by current legislations on data privacy (i.e., healthcare data, religion, sexuality and location data); we simulated a set of messages to be published (and protected) referring to those entities by taking their descriptions in their corresponding Wikipedia articles. Note that, due to the highly informative nature of Wikipedia articles describing the entities to be protected, using this text as messages to be published represents a very challenging test bed from the perspective of document sanitization [33, 34].

As a benchmark for assessing the accuracy of our proposal regarding both sensitive term detection and term disambiguation, the collaboration of a human expert was required. More specifically, for the first operation, the human expert was asked to manually identify the textual terms in the simulated messages that, according to her opinion, unequivocally disclosed the underlying entity to be protected with respect to the level of disclosure specified by the user (i.e., *AL*); regarding the second operation, the human expert was asked to manually validate the terms that were correctly disambiguated by the system with respect to the senses available in WordNet for such terms. According to that judgement, the accuracy of the process for detecting sensitive term was quantified in terms of *precision*, *recall* and *f-measure*; while the accuracy of the semantic disambiguation process was quantified just in terms of *precision*.

*Precision*, in the first operation, it measures the percentage of automatically identified terms in the message (*S*) that are truly sensitive according to the expert's opinion (*H*). A high precision is desirable, because it indicates that the system has incurred in a low number of false positives, which may unnecessarily hamper the utility and readability of the protected messages. See equation (4) for a formal representation of *precision*:

$$Precision = \frac{|S \cap H|}{|S|} \times 100 \qquad (4)$$

In the second operation, *precision* is just the percentage of properly disambiguated terms, according to the expert's opinion. In this case, a high *precision* is desirable in order to replace sensitive terms by semantically coherent generalizations.

On the other hand, *recall*, which only applies to the first operation, measures the percentage of sensitive terms correctly detected by the system (*S* ∩ *H*) from the total number of the terms detected by the human expert (*H*). A high *recall* is desirable because it indicates that the protected message fulfils with the privacy requirements of the user. See equation (5) for a formal representation of *recall*:

$$Recall = \frac{|S \cap H|}{|H|} \times 100 \qquad (5)$$

Finally, *f-measure* provides the harmonic mean of *precision* and *recall* and, thus, summarizes the accuracy of the process in charge of detecting sensitive terms in the messages to be published. See equation (6) for formal representation of *f-measure*:

$$F-measure = \frac{2 \times Recall \times Precision}{Recall + Precision} \qquad (6)$$

Evaluation results that show the accuracy of the sensitive term detection process are depicted in Table 5. To evaluate the effect of the configuration of the privacy requirements, each entity has been protected and evaluated (by the human expert) for two access levels (*AL*) with different degrees of generality.

| Entity/ Wikipedia article | Related *ST* | # words in text | # noun phrases | Access Level | H | S | Recall | Precision | F-measure |
|---|---|---|---|---|---|---|---|---|---|
| HIV | Health | 49 | 20 | Condition | 9 | 7 | 77.77 % | 100 % | 87 % |
| | | | | Infection | 4 | 3 | 75 % | 100 % | 85 % |
| Christianity | Religion | 66 | 22 | Belief | 10 | 9 | 90 % | 100 % | 94 % |
| | | | | Religion | 7 | 5 | 71 % | 100 % | 83 % |
| Homosexuality | Sexual orientation | 78 | 26 | Process | 6 | 6 | 100 % | 100 % | 100 % |
| | | | | Sexual activity | 5 | 4 | 80 % | 100 % | 88 % |
| Berlin | Census data | 105 | 31 | Location | 10 | 8 | 80 % | 100 % | 88 % |
| | | | | City | 3 | 3 | 100 % | 100 % | 100 % |

**Table 5.** Evaluation results for the process in charge of detecting sensitive terms

In all cases, the system achieves perfect *precision* because, as stated in the privacy requirements, it sanitizes terms that are semantic specializations of the entities defined as access levels; thus, by definition, all the detected terms that may disclose the entity must be protected. On the other hand, *recall* figures fluctuated between 71-100%, showing that there is still room for improvement. Indeed, according to the expert assessment, some *combinations* of terms that are not actual specializations of the entity to be protected, but that can be related on some way with it, may also enable disclosure and should be adequately protected. For example, an informed attacker may infer that a publisher suffers from a certain sensitive disease from the fact that specific treatments or symptoms are mentioned in a discourse, despite of the fact that the disease has been already sanitized in the published message and that those terms are not specializations of the former. We are currently working on this issue and we provide some insights on how to tackle it in the next section. Finally, we can also see that the recall (i.e., the accuracy of the privacy protection) tends to increase as more general terms are defined as *ALs*. Indeed, a more general *AL* will impose a stronger restriction and force the system to sanitize more terms and, thus, the outcome would tend to offer a more robust protection.

Regarding the accuracy achieved by the semantic disambiguation process, Table 6 shows the evaluation results that have been obtained. As it has been previously explained, in this case the human expert just validates the terms that have been correctly disambiguated by the proposed system according to the senses available for such terms in WordNet. Results reflect that, on average, the scheme disambiguated 66% of the terms correctly; even though this value may seem on the low side, it is coherent with the state of the art in semantic disambiguation [24], which rarely achieves very high precision figures. Moreover, improperly disambiguated terms would only affect the semantic coherence of the protected message, but not the privacy of the user, which is our main goal.

| Entity/ Wikipedia article | Related *ST* | Access Level | H | S | Precision |
|---|---|---|---|---|---|
| HIV | Health | Condition | 7 | 4 | 57 % |
| | | Infection | 3 | 2 | 66 % |
| Christianity | Religion | Belief | 9 | 5 | 55 % |
| | | Religion | 5 | 3 | 60 % |
| Homosexuality | Sexual orientation | Process | 6 | 4 | 66 % |
| | | Sexual activity | 4 | 2 | 50 % |
| Berlin | Census data | Location | 8 | 6 | 75 % |
| | | City | 3 | 3 | 100 % |

**Table 6.** Evaluation results for the process in charge of disambiguating terms

## 6. Conclusions and future work

In this paper, we proposed a privacy-preserving content-driven access control mechanism for textual publications in social networks. Contrary to related works [10] [11, 12], the proposal is content driven in the sense that the semantics of the messages are automatically assessed in order to detect the sensitive information they contain according to the privacy requirements of the publishers. These requirements are defined in general (i.e., an allowed level of disclosure is defined for the different contact types defined in the social network), and the publications whose contents are related to these requirements are automatically protected. To do so, the sensitive information is sanitized and different versions of the publication are generated according to the access level of the readers. Thus, the privacy enforcement is transparent both to the publishers and readers, thus requiring no administrative efforts at the publication time, contrary to most related works [11, 12]. In addition, the proposed mechanism is flexible enough to be incorporated in any social network that publishes messages and classifies contacts into categories.

As future work, we plan to develop a functional implementation in a real OSN in order to a survey a group of social network users on the usability and utility of the proposed system. For this purpose, we will engineer the privacy requirements to be considered within the scope of the network. Furthermore, in order to alleviate users from completely specifying their privacy requirements, we will also consider the automatic inference of access control rules according to the social relationships implemented in the social network (e.g., the privacy rules for friends could be same for the friends of friend). At this respect, a machine learning approach [35] can also be considered to semi-automatize the configuration of privacy rules. Finally, as it has been highlighted in the evaluation section, the user's privacy can also be compromised by the (co-)occurrence of information that is correlated to the sensitive topic to be protected. We are currently working on automatic solutions to address this issue that, in a nutshell, would assess the disclosure that potentially correlated terms may produce for a sensitive one according to their mutual information, which is computed from the information distribution of data in large corpora [33, 34]. We plan to incorporate them to the developed system in the near future in order to improve the assessment of privacy risks by detecting correlated terms or term aggregations that may disclose more information about a sensitive topic than the one specified in the privacy rules.

## Acknowledgements and disclaimer

Authors are solely responsible for the views expressed in this paper, which do not necessarily reflect the position of UNESCO nor commit that organization. This work was partly supported by the European Commission under the H2020 project CLARUS, by the Spanish Ministry of Science and Innovation (through projects CO-PRIVACY TIN2011-27076-C03-01, ICWT TIN2012-327570 and SmartGlacis TIN2014-57364-C2-1-R) and by the Government of Catalonia (under grant 2014 SGR 537). This work was also made possible through the support of a grant from Templeton World Charity Foundation (TWCF0095/AB60 CO-UTILITY). The opinions expressed in this paper are those of the authors and do not necessarily reflect the views of Templeton World Charity Foundation.## References